\pdfoutput=1
\documentclass{optica-article}

\journal{opticajournal}

\articletype{Research Article}

\usepackage{amsmath,booktabs}
\newcommand{\dd}{\mathrm{d}}

\begin{document}

\title{Edge photocurrent distribution in luminescent solar concentrators: analytical treatment and solar cell arrangement}

\author{Ilya Sychugov\authormark{*}}

\address{Department of Applied Physics, KTH -- Royal Institute of Technology, Stockholm 11419, Sweden}

\email{\authormark{*}ilyas@kth.se}

\begin{abstract*}
In luminescent solar concentrators (LSCs), the light flux along the edges is non-uniform, which causes current mismatch between series-connected edge solar cells. Here, the optical path distribution of photons arriving at an edge point was derived analytically for a rectangular LSC and generalized to convex polygons, with reabsorption and scattering included as a geometric series. The model agrees with Monte Carlo simulations and reproduces measured edge cell currents of a $60\times60$\,cm$^2$ LSC within 2--3\%. In a square LSC, the series mismatch reduces the current by 8--12\% for four cells per edge. Rules for edge solar cell arrangement are formulated.
\end{abstract*}

\section{Introduction}
Luminescent solar concentrators (LSCs) were introduced in the 1970s to concentrate sunlight by means of photoluminescence (PL) in a planar waveguide \cite{Debije2012}. Fluorophores embedded in a glass or polymer slab absorb a part of the incident light, both direct and diffuse, and a significant fraction of the emitted PL ($\sim$75\% for a refractive index of 1.5) is trapped by total internal reflection until it reaches the edges, where it is converted to electricity by solar cells. Recently, LSCs became attractive again as semitransparent building-integrated photovoltaic (BIPV) units, commonly described as ``solar windows'' \cite{Meinardi2017NRM,Richards2023}. Indeed, ground-mounted photovoltaics competes with other land use, while in cities the rooftop area is too small to supply tall buildings \cite{Meinardi2017NRM,BorjaBlock2024}. At the same time, nearly zero-energy and positive-energy buildings are promoted by European regulations \cite{Magrini2020,Meinardi2024}. Large glazed facades thus become a natural target for semitransparent, color-neutral photovoltaics \cite{Traverse2017,Li2023NRM}, where solar cells of an LSC can be hidden in the window frame.

\begin{figure}[htbp]
\centering\includegraphics[width=\linewidth]{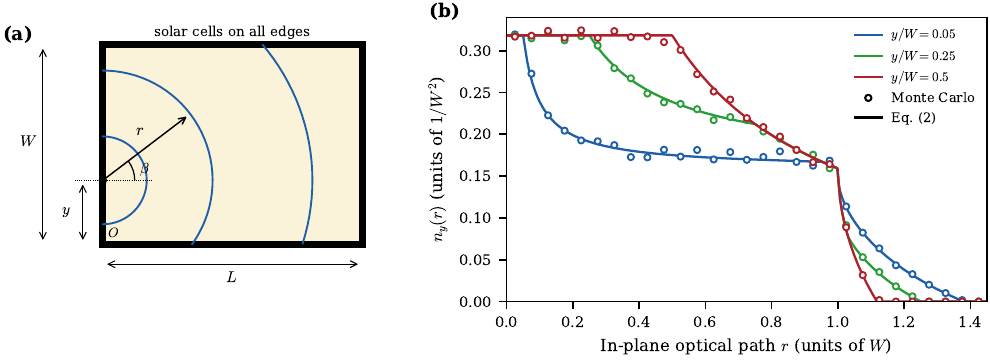}
\caption{(a) Geometry of a rectangular LSC with solar cells on all four edges (thick lines), length $L$ (perpendicular to the considered edge) and width $W$ (along it). The considered edge is at $x=0$, and a point on this edge is at a distance $y$ from the corner $O$. Photons arriving at this point at an in-plane angle $\beta$ from the edge normal originate from arcs of radius $r$ inside the device. (b) Optical path distribution $n_y(r)$ per unit total emission for photons arriving at three points of the edge of a square LSC ($L=W$). Lines are from Eq.~(\ref{eq:ny}), and symbols are from Monte Carlo simulations.}
\label{fig:paths}
\end{figure}

Over the years, LSC development has moved from small laboratory samples to large-area devices \cite{Li2016,Zhou2026}, laminated and certification-grade glazing units \cite{Bergren2018,Meinardi2024,Huang2022}, and pilot installations \cite{Aste2019}, with standardized reporting protocols \cite{Yang2022}. Realistic efficiency targets for commercial-scale modules were estimated as 9.0\% for opaque and 4.5\% for semitransparent (50\% average visible transmission) devices with ideal quantum dots \cite{Richards2023}, and as $\sim$1\% for window-appropriate transparency with existing luminophores \cite{deBruin2024}. For power windows, figures of merit beyond the efficiency are also important \cite{Needell2021}.

Historically, the concentration ratio was the main figure of merit for an LSC, and solar cells were often attached to one edge only. For solar windows, instead, the full perimeter is covered, and the edge cells are connected in series to obtain a useful voltage. In BIPV installations, current mismatch between series-connected cells is a known problem, with losses of 5--25\% depending on the electrical configuration \cite{Ravyts2019}. In an LSC, the light flux along an edge is not uniform, with a maximum at the edge center. This was measured early \cite{Heidler1981,Decardona1985}, and its consequence for series-connected cells was pointed out at the same time \cite{Decardona1985}. The effect of shape and cell placement was analyzed by a method of images \cite{Loh1986}. Wilson et al. demonstrated that shortening of the central cells along an edge of a $60\times60$\,cm$^2$ LSC allows one more cell in series and increases the output power by 15\% \cite{Wilson2017}. In that work, the cell lengths were obtained from a polynomial fit of a ray-traced edge profile. Alternatively, the mismatch can be avoided by guiding light to a single edge \cite{Han2022}. Edge-resolved currents were also used to characterize large-area LSCs \cite{Li2020}, and position-dependent photocurrents were observed for long series-connected cells under point excitation \cite{Zhou2022}.

We have previously shown that the waveguiding efficiency of an LSC can be analytically derived from the distribution of optical path lengths of isotropic emitters \cite{Sychugov2019,Sychugov2020}. This description compares well with Monte Carlo ray tracing \cite{Zheng2022,deBruin2023}, while the results are obtained instantly. Related analytical treatments describe reabsorption in circular devices \cite{tenKate2014}, successive generations of re-emitted light \cite{Merkx2017,Matsumura2020}, and concentration limits \cite{Klimov2016}. These models, however, give the total flux reaching the perimeter. Fujieda et al. introduced an edge collection function, a probability for a photon emitted at a given spot to be collected by a given edge \cite{Fujieda2019,Fujieda2020JPE}, and applied it to energy-harvesting displays with edge photodiodes \cite{Yunoki2019}. This function is resolved in the emitter position rather than in the position along the edge, and scattering and re-emission are not included.

Here, the edge flux profile is derived from the receiver point of view. For a fixed point on the edge, the probability density of the optical path of arriving photons is obtained first, and all losses and recycling processes are then included through this distribution. The results were verified by Monte Carlo simulations and by published edge-current measurements, and explicit rules for the arrangement of edge solar cells were obtained. The treatment is presented for a rectangle and generalized in Supplement 1 to other convex shapes, including triangular and hexagonal devices. Derivations, numerical methods and the full set of simulation data are also given there.

\section{Theory}
\subsection{Optical path distribution at an edge point}
We consider a rectangular LSC of area $A=LW$ with solar cells covering all four edges, so that every photon reaching the perimeter is absorbed [Fig.~\ref{fig:paths}(a)]. This full edge coverage is the most relevant case for solar windows, while mirrors and partial coverage are considered in Section~\ref{sec:mirrors}. The edge under consideration is along $x=0$ (length $W$), and the opposite edge is at $x=L$. A point on this edge is at a distance $y$ from the corner $O$ and $W-y$ from the other corner. In the plane of the slab, an isotropic emitter at a distance $r$, seen at an angle $\beta$ from the edge normal, sends a fraction $\cos\beta\,\dd y/(2\pi r)$ of its trapped emission to the edge element $\dd y$. For uniformly distributed emitters the factor $r$ cancels out with the area element in polar coordinates, and the photon flux per unit edge length becomes (see Section~\ref{S-s:ny} in Supplement 1)
\begin{equation}
j(y)=\int_0^\infty n_y(r)\,e^{-\alpha k r}\,\dd r,
\label{eq:jdef}
\end{equation}
\begin{equation}
n_y(r)=\frac{g_y(r)+g_{W-y}(r)}{2\pi A},
\label{eq:ny}
\end{equation}
where $g_s(r)$ is the range of $\sin\beta$ within which the point at a distance $r$ lies inside the device on the side of the corner at a distance $s$ ($s=y$ or $W-y$). It is limited from above by the side edge ($\sin\beta<s/r$) and, for $r>L$, from below by the opposite edge ($\sin\beta>\sqrt{1-L^2/r^2}$), which gives an elementary piece-wise function of $r$ [Eqs.~(\ref{S-eq:gs1}) and (\ref{S-eq:gs2}) in Supplement 1]. Here, $\alpha$ is the total attenuation coefficient and $k=\psi_m/\sin\psi_m\approx1.13$ is the mean out-of-plane path extension of trapped photons for $n=1.5$ (Section~\ref{S-s:k} in Supplement 1) \cite{Soti1996}. Note that this value is slightly lower than $k\approx1.14$ used previously \cite{Sychugov2019}, which corresponds to a uniform weighting of the elevation angles. The function $n_y(r)$ is the probability density of the in-plane optical path for photons arriving at point $y$. It is constant, $1/(\pi A)$, for $r<\min(y,W-y,L)$, where the whole half-circle lies inside the device, and decreases for larger $r$ as the half-circle is truncated by the device edges [Fig.~\ref{fig:paths}(a)]. Summed over all edge points, it gives the path distribution $p(r)$ of the whole device \cite{Sychugov2019} [Eq.~(\ref{S-eq:sum}) in Supplement 1]. Numerical simulations indeed converge to Eq.~(\ref{eq:ny}) [Fig.~\ref{fig:paths}(b)].

\subsection{Edge flux profile}
Now, when the path distribution at an edge point is established, the edge flux profile follows directly from Eq.~(\ref{eq:jdef}). Without attenuation, it is
\begin{multline}
j_0(y)=\frac{1}{2\pi A}\Big[L\Big(\arctan\frac{y}{L}+\arctan\frac{W-y}{L}\Big)\\
+\frac{y}{2}\ln\Big(1+\frac{L^2}{y^2}\Big)+\frac{W-y}{2}\ln\Big(1+\frac{L^2}{(W-y)^2}\Big)\Big],
\label{eq:j0}
\end{multline}
and for a square the corner value is 65\% of the center value, as first obtained by Goetzberger and Greubel \cite{Goetzberger1977}. With attenuation one obtains (Section~\ref{S-s:profile} in Supplement 1)
\begin{equation}
j(y)=\frac{1}{2\pi A\alpha k}\Big[2-I_L-I_y-I_{W-y}\Big],
\label{eq:jatt}
\end{equation}
where $I_L$ is the contribution of rays ending at the opposite edge (an incomplete Bickley function), and the side-edge terms are expressed through the exponential integral $E_1(x)=\int_x^\infty e^{-t}/t\,\dd t$:
\begin{align}
I_s&=G_s(1)-G_s\big(s/\sqrt{s^2+L^2}\big),\label{eq:Is}\\
G_s(c)&=c\,e^{-\alpha ks/c}-\alpha ks\,E_1(\alpha ks/c),\nonumber
\end{align}
with $s=y$ or $W-y$. For large devices ($\alpha L,\alpha W\gg1$) the profile becomes flat, $j\to1/(\pi A\alpha k)$, except within $\sim1/\alpha k$ from the corners, where it drops to one half of this value. Thus, the corner-to-center contrast is non-monotonic with the device size.

\begin{figure}[htbp]
\centering\includegraphics[width=\linewidth]{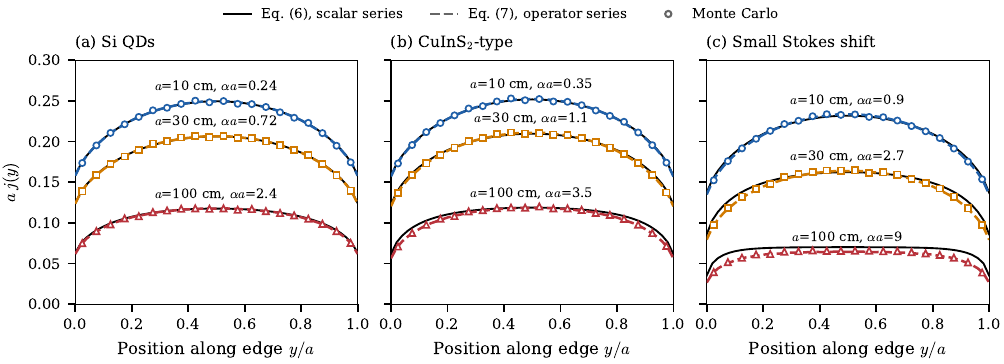}
\caption{Edge photon flux profiles (per unit edge length, per trapped first emission, multiplied by $a$) for square LSCs with side $a$=10, 30 and 100\,cm and the three fluorophores of Table~\ref{S-tab:fl} in Supplement 1. Lines are from the scalar series, Eq.~(\ref{eq:scalar}) (solid black), and from the operator series, Eq.~(\ref{eq:operator}) (dashed). Symbols are Monte Carlo results ($1.5\times10^6$ photons). The corresponding waveguiding efficiencies are listed in Table~\ref{S-tab:eff} in Supplement 1.}
\label{fig:prof}
\end{figure}

\subsection{Scattering and reabsorption}
In practice, a part of the attenuated photons is not lost but re-emitted, which is important for strongly reabsorbing fluorophores and large devices. At each scattering or reabsorption event a photon is re-emitted into a waveguided mode with a probability $p=\delta(\alpha_{sc}+QY\,\alpha_{re})/\alpha$, where $\alpha=\alpha_{sc}+\alpha_{re}+\alpha_{mx}$ and $\delta=\cos\theta_c\approx0.75$. If the secondary emitters of each generation are uniformly distributed, every generation gives the same profile, attenuated by $p[1-f(\alpha)]$, where $f$ is the waveguiding efficiency \cite{Sychugov2019}. The geometric series then becomes
\begin{equation}
j_\mathrm{tot}(y)=\frac{j(y)}{1-p\,(1-f)},
\label{eq:scalar}
\end{equation}
i.e., the profile shape remains the same. In reality, secondary emitters are depleted near the edges, since photons emitted there are collected before they interact. An exact result follows from the operator form of the same series (Section~\ref{S-s:series} in Supplement 1):
\begin{equation}
j_\mathrm{tot}(y)=\int_A \Phi_y(P)\,S(P)\,\dd A,\quad S=s_0+pKS,
\label{eq:operator}
\end{equation}
where $s_0=1/A$, $\Phi_y(P)=\cos\beta\,e^{-\alpha kr}/(2\pi r)$ is the edge response to a unit emitter at point $P$, and $K(P,P')=\alpha k\,e^{-\alpha k|P-P'|}/(2\pi|P-P'|)$ is the density of interaction events at $P$ per photon emitted at $P'$. It is clear that for a given edge point all photon histories enter only through the position of the last emission event. Equation~(\ref{eq:scalar}) is recovered when $K$ is replaced by its average over a uniform source, $1-f$. Equation~(\ref{eq:operator}) can be considered as a receiver-resolved counterpart of multiple-generation transport models \cite{Merkx2017,Matsumura2020}, while integration of $\Phi_y$ along an edge gives the edge collection function \cite{Fujieda2019,Fujieda2020JPE} for a single wavelength. After discretization of the device area, Eq.~(\ref{eq:operator}) becomes a linear system, which is solved directly without photon sampling. This recurrent solution is fast, although not instant as the closed-form expressions.

\section{Verification}
\subsection{Monte Carlo simulations}
To verify the obtained expressions, they were first compared with numerical simulations. Three representative fluorophores were considered (Table~\ref{S-tab:fl} in Supplement 1): Si quantum dots (QDs) in a glass/polymer laminate with a large Stokes shift and low reabsorption \cite{Zhou2026}, a CuInS$_2$-type emitter with a moderate Stokes shift and high $QY$, and a small-Stokes-shift emitter of the CsPbBr$_3$/CdSe type with strong reabsorption and $QY\approx1$. The Monte Carlo code includes isotropic three-dimensional emission, escape cones, scattering and reabsorption events (Section~\ref{S-s:mc} in Supplement 1). Edge profiles for square LSCs of side $a$ ($L=W=a$) are shown in Fig.~\ref{fig:prof}. For Si QDs, the scalar series [Eq.~(\ref{eq:scalar})] reproduces both the profile shape and the total efficiency within 0.3\% for all sizes. For the small-Stokes-shift emitter at $a=100$\,cm, where half of the collected photons were re-emitted at least once, it overestimates the efficiency by 11\% (0.267 vs. 0.240), and the profile is too flat. The operator series [Eq.~(\ref{eq:operator})] gives 0.235 and follows the simulated profile. In all cases the operator series is within 1--2\% of the Monte Carlo results, where the remaining difference stems from the approximation $l\approx kr$ and the source discretization.

\begin{figure}[htbp]
\centering\includegraphics[width=\linewidth]{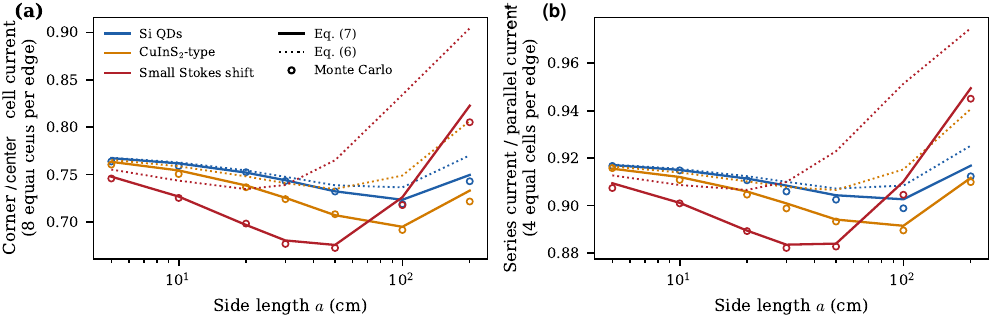}
\caption{Size dependence of (a) the ratio of corner to center cell currents for eight equal cells per edge and (b) the current of four equal series-connected cells relative to their parallel connection. Solid lines: Eq.~(\ref{eq:operator}); dotted lines: Eq.~(\ref{eq:scalar}); symbols: Monte Carlo simulations.}
\label{fig:size}
\end{figure}

For practical purposes, the profile integrated over each cell is the relevant quantity. In Fig.~\ref{fig:size}, the ratio of the corner to the center cell current for eight equal cells per edge and the current of four series-connected equal cells relative to their parallel connection are shown. The operator series agrees with Monte Carlo simulations within 0.01 for all fluorophores and sizes of 5--200\,cm. Central cells receive 1.3--1.5 times more current than corner cells in most of this range, and the series connection of four equal cells reduces the current by 8--12\%. Both ratios are non-monotonic with the device size. Weak attenuation suppresses mainly the long optical paths, which dominate the signal near the corners, while for $\alpha a\gtrsim3$ the corner dip of the profile, with a width of $\sim1/\alpha k$, becomes narrower than a cell, and the cell currents become more uniform.

\subsection{Comparison with experiments}
\begin{figure}[htbp]
\centering\includegraphics[width=\linewidth]{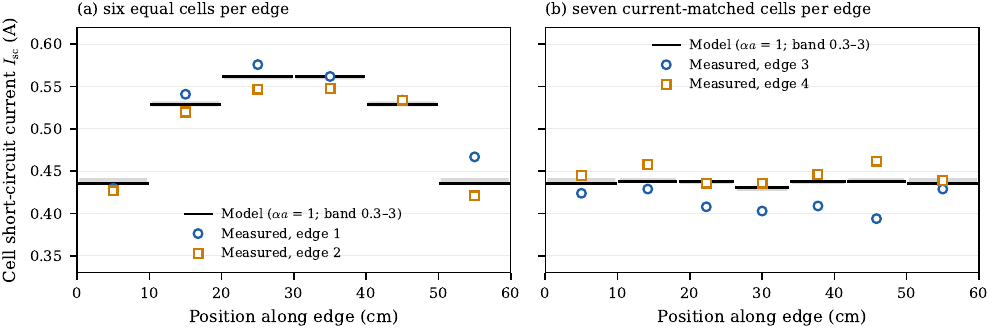}
\caption{Short-circuit currents of edge cells of a $60\times60\times0.3$\,cm$^3$ LSC measured by Wilson et al. \cite{Wilson2017} (symbols) and calculated from Eq.~(\ref{eq:jatt}) integrated over each cell (black lines, $\alpha a=1$; gray bands, $\alpha a=0.3$--3). (a) Six equal 10\,cm cells per edge. (b) Seven current-matched cells per edge. One scale factor was fitted to the data in (a), and the same factor was used in (b).}
\label{fig:exp}
\end{figure}

While the simulations test the derivation, a comparison with measurements tests also the model assumptions, such as ideal edge coverage and a uniform device. The most detailed experimental data available are the short-circuit currents of 26 c-Si strip cells attached to the edges of a $60\times60\times0.3$\,cm$^3$ PMMA LSC with 400\,ppm Lumogen F Rot 305 \cite{Wilson2017}. Two edges contain six equal 10\,cm cells, and two other edges contain seven current-matched cells of 10.00, 8.35, 7.85, 7.60, 7.85, 8.35 and 10.00\,cm length. The measured currents together with Eq.~(\ref{eq:jatt}) integrated over each cell are shown in Fig.~\ref{fig:exp}. The tail absorption of this dye in PMMA gives $\alpha a\approx0.5$--2.5 for the waveguided PL, and the ray-traced profile of this device is best reproduced with $\alpha a\approx1$ \cite{WilsonThesis}. The model was therefore evaluated for $\alpha a=1$, while the range $\alpha a=0.3$--3 changes the relative currents by less than 1\%. One scale factor was fitted to the twelve cells of the equal-cell edges, and the matched edges were then calculated without adjustable parameters. For equal cells, the measured profile is reproduced with a root-mean-square deviation of 2--3\%, and the end-to-center current ratio (0.775) is close to the measured 0.75--0.77. For the matched cells, the model predicts a spread below 2\%. Edge 4 follows this prediction within 3\%, while edge 3 is on average 5\% lower with a larger spread. This was attributed by the authors to a thickness variation of the cast sheet and to partial shading of the cells by their busbars \cite{Wilson2017}, which are not included in the model. The equal-current lengths calculated from Eq.~(\ref{eq:jatt}) for seven cells, 9.9--10.1, 8.2--8.4, 7.8--7.9 and 7.7--7.8\,cm from the corner to the center ($\alpha a=0.3$--3), agree within 0.15\,cm with the ray-tracing results of Ref.~\cite{Wilson2017} (10.15, 8.25, 7.77 and 7.66\,cm).

Heidler measured the profile along the output edge of a right isosceles triangle LSC with reflectors at the two short sides, and the ratio of the edge-average to the edge-center intensity was 0.935 \cite{Heidler1981}. With specular reflectors, this geometry unfolds into a square with four absorbing edges, for which Eq.~(\ref{eq:jatt}) gives 0.90--0.91 ($\alpha a=0$--3). Diffuse reflectors used in the experiment redistribute the light along the edge, making the profile slightly flatter. Sidrach de Cardona et al. moved a 2\,cm cell along the edge of a 50\,cm square LSC and measured a decrease of the current to 0.87 of its maximum at the corner \cite{Decardona1985}, while Eq.~(\ref{eq:jatt}) gives 0.63--0.70 for $\alpha a\le10$. In that experiment, the rest of the edge was not covered by solar cells, and a reflecting edge made the decrease less pronounced than a blackened one \cite{Decardona1985}. The light reflected from uncovered edge parts also increases the signal of a single cell, which was corrected by an empirical factor of 1.28 in Ref.~\cite{Heidler1981}. Thus, single movable cell measurements underestimate the non-uniformity for a fully covered edge, and the present model can provide the corresponding correction analytically.

The model can also be used to interpret size-dependent efficiencies, which are often measured with series-connected edge cells. For example, glass laminated LSCs of $7.8\times7.8$ to $31.2\times31.2$\,cm$^2$ and $61\times122$\,cm$^2$ with series-connected c-Si cells showed a decrease of the power conversion efficiency from 2.80\% to 0.33\% \cite{Zhang2015}. For the square devices with one or two equal cells per edge, all cells receive the same current due to symmetry, and no mismatch occurs. For the $61\times122$\,cm$^2$ device with seven and three 15.6\,cm cells along the long and short edges, respectively, Eq.~(\ref{eq:jatt}) gives a string current of 0.87--0.93 of the average cell current ($\alpha a_\mathrm{eq}=0.3$--3, where $a_\mathrm{eq}$ is the side of a square of equal area). Here, the cells were assumed to be centered on each edge, since their exact positions were not reported. Thus, the series mismatch is expected to contribute 7--13\% to the efficiency decrease of the largest device, in addition to the waveguiding losses.

\section{Mirrors and partial edge coverage}\label{sec:mirrors}
\begin{figure}[htbp]
\centering\includegraphics[width=\linewidth]{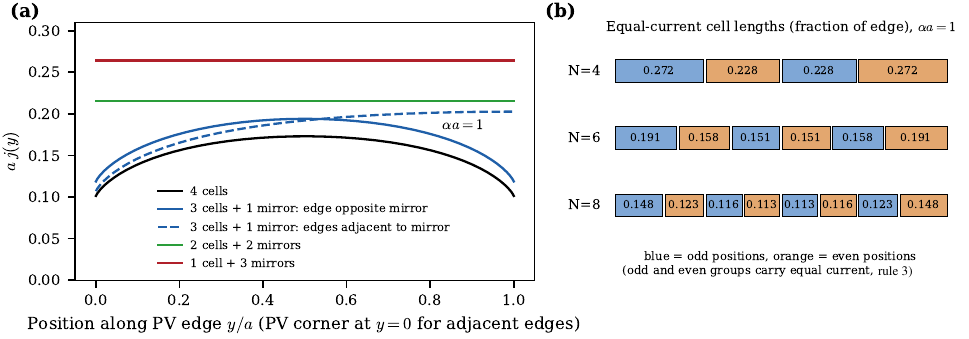}
\caption{(a) Edge profiles for different combinations of solar cell and mirror edges at $\alpha a=1$ (square, scalar series). (b) Equal-current cell lengths (fraction of the edge) for $N$=4, 6 and 8 cells per edge at $\alpha a=1$. Odd (blue) and even (orange) positions form the two groups of the interleaved arrangement (rule 3).}
\label{fig:design}
\end{figure}

So far, all device edges were covered by solar cells. In practice, it is often important to reduce the number of cells, and a part of the edges is then covered by mirrors. Specular mirrors can be treated by unfolding the device across them \cite{Lu2023} (Section~\ref{S-s:mirrors} in Supplement 1). For two opposite solar cell edges and two mirrors, the unfolded device is an infinite strip of width $L$, and the edge profile is exactly uniform, $j=f_\mathrm{s}/2W$. Here, $f_\mathrm{s}=2[1-\mathrm{Ki}_2(\alpha kL)]/(\pi\alpha kL)$ is the exact strip efficiency, and $\mathrm{Ki}_2$ is a Bickley function. One solar cell edge with three mirrors corresponds to a strip of width $2L$ and also gives a uniform profile. Three solar cell edges with one mirror unfold into a $2L\times W$ rectangle with four absorbing edges, so that Eqs.~(\ref{eq:jatt})--(\ref{eq:operator}) can be directly applied. Along the edges adjacent to the mirror, the flux decreases only towards the corner shared with another solar cell edge [Fig.~\ref{fig:design}(a)]. When solar cells cover only a small part of an otherwise reflecting perimeter, the trapped light approaches a uniform photon gas, and the collected flux does not depend on the cell position \cite{Song2018}.

\section{Arrangement of edge solar cells}
Now, when the edge profiles are established for different edge configurations, we can turn to the arrangement of edge solar cells. For an absorbed photon flux $N_\mathrm{abs}$ (s$^{-1}$, whole device), the photocurrent generated per unit edge length is
\begin{equation}
i(y)=e\,N_\mathrm{abs}\,QY\,\delta\,\mathrm{EQE}_\mathrm{cell}\,j_\mathrm{tot}(y),
\label{eq:current}
\end{equation}
assuming that the cell covers the full edge thickness. Here, $\mathrm{EQE}_\mathrm{cell}$ is taken as angle-independent. Waveguided photons reach the edge at elevation angles up to $\psi_m\approx48^\circ$, and the angular distribution of the edge emission is affected by reabsorption \cite{Parel2015}. This gives a small correction which depends only weakly on the position along the edge. A cell between $y_1$ and $y_2$ gives $\int_{y_1}^{y_2}i(y)\,\dd y$. In practice, a substantial voltage is required for applications, while the voltage of a single cell is low, e.g., $\sim$0.6\,V for c-Si. Therefore, the edge cells are often connected in series, along each edge and around the whole perimeter \cite{Wilson2017}. In a series string the current is limited by the smallest of the cell currents. In contrast, for parallel connection the currents add up, and the non-uniformity causes no losses, but the voltage remains that of a single cell. Note that junctions connected in series within one multi-junction cell along the edge behave as a series string. From the results above, the following rules can be formulated for series-connected cells in square and rectangular LSCs with full edge coverage.

\textit{1. Series connection of equal cells.} The current of a string of equal cells along one edge relative to parallel connection is 0.95 for 3 cells, 0.91 for 4, 0.86 for 6, and 0.82--0.85 for 8 cells ($\alpha a\le3$, Table~\ref{tab:design}). For large devices ($\alpha a\approx10$) the loss is smaller, 0.96 for 4 and 0.89 for 8 cells, since the profile becomes flatter. With strong recycling the loss increases by up to 2--3 percentage points [Fig.~\ref{fig:size}(b)]. Variations between edges, such as differences in thickness or cell alignment, add to the mismatch of a string around the whole perimeter \cite{Wilson2017}.

\textit{2. Equal-current cell lengths.} The mismatch is removed if the corner cells are longer (Table~\ref{tab:design}, Fig.~\ref{fig:design}(b)). These lengths are nearly independent of the device size for $\alpha a\le3$. For four cells per edge they are 0.27 and 0.23 of the edge length, i.e., corner cells are $\sim$17\% longer. For eight cells they are 0.147, 0.123, 0.116 and 0.113 from the corner to the center. Strong recycling changes these values by less than 0.005 of the edge length. Current matching by cell length was experimentally demonstrated with a 15\% increase of the output power due to one more cell per edge \cite{Wilson2017} (Fig.~\ref{fig:exp}).

\textit{3. Interleaved series--parallel connection.} For an even number of \emph{equal} cells per edge, cells $i$ and $N+1-i$ receive the same current due to symmetry and have opposite parity. Thus, all odd-position cells connected in parallel and all even-position cells connected in parallel form two blocks with \emph{exactly} equal currents, which can be connected in series. For a square LSC, four edges give eight identical blocks without mismatch losses (Section~\ref{S-s:rect} in Supplement 1).

\textit{4. Rectangular devices.} The current per unit length is only slightly different for long and short edges (ratio of 0.92--1.07 for aspect ratio 2 and 0.84--1.13 for 4). Series-connected blocks from different edges should then contain total cell lengths proportional to the edge lengths (Section~\ref{S-s:rect} in Supplement 1). For other regular shapes of equal area, the total efficiency is similar \cite{Loh1986,Sychugov2020}, while the edge uniformity improves with the number of sides: the series loss for four equal cells per edge is 15--17\% for a triangle, 8--9\% for a square and 3--4\% for a hexagon, while a circle gives a uniform profile (Section~\ref{S-s:shapes} in Supplement 1).

\textit{5. Choice of model.} For large-Stokes-shift fluorophores, where the recycled fraction is below $\sim$10--20\%, Eqs.~(\ref{eq:jatt}) and (\ref{eq:scalar}) are sufficient. For devices with strong reabsorption or scattering, Eq.~(\ref{eq:operator}) should be used.

\begin{table}[htbp]
\centering\small
\caption{Equal-current cell lengths along one edge of a square LSC (fraction of the edge length from the corner to the center, symmetric) and the current of a series string of equal cells relative to parallel connection.}
\label{tab:design}
\setlength{\tabcolsep}{3pt}
\begin{tabular}{cccc}
\toprule
$N$ & $\alpha a$ & equal-current lengths & equal cells\\
\midrule
3 & 0.3 & 0.348, 0.303 & 0.951\\
  & 1   & 0.349, 0.301 & 0.948\\
  & 3   & 0.348, 0.303 & 0.951\\
  & 10  & 0.341, 0.319 & 0.977\\
\midrule
4 & 0.3 & 0.270, 0.230 & 0.915\\
  & 1   & 0.272, 0.228 & 0.908\\
  & 3   & 0.271, 0.229 & 0.912\\
  & 10  & 0.261, 0.239 & 0.956\\
\midrule
6 & 0.3 & 0.189, 0.159, 0.152 & 0.868\\
  & 1   & 0.191, 0.158, 0.151 & 0.856\\
  & 3   & 0.191, 0.158, 0.152 & 0.856\\
  & 10  & 0.180, 0.160, 0.159 & 0.917\\
\midrule
8 & 0.3 & 0.146, 0.124, 0.116, 0.113 & 0.839\\
  & 1   & 0.148, 0.123, 0.116, 0.113 & 0.823\\
  & 3   & 0.149, 0.122, 0.116, 0.113 & 0.817\\
  & 10  & 0.139, 0.121, 0.120, 0.120 & 0.885\\
\bottomrule
\end{tabular}
\end{table}

\section{Conclusion}
To conclude, an analytical description of the photocurrent distribution along the edges of a rectangular LSC has been developed and generalized to arbitrary convex polygons. The probability density of the optical paths for photons arriving at a given edge point was derived in closed form, and the edge profile was expressed through elementary functions, the exponential integral and one smooth single integral. Scattering and reabsorption were included as a geometric series. Its scalar form is accurate for large-Stokes-shift fluorophores, while the operator form agrees with Monte Carlo simulations also for strongly reabsorbing emitters. The model reproduces the measured edge cell currents of a $60\times60$\,cm$^2$ LSC within 2--3\%, and the current-matched cell lengths within 0.15\,cm. In a square LSC, central cells receive 1.3--1.5 times more current than corner cells, which reduces the current by 8--12\% for four and by up to $\sim$18\% for eight equal series-connected cells per edge. Among regular shapes, the non-uniformity is strongest for a triangle and weakest for a hexagon, and it vanishes for a circle. This loss can be avoided by parallel connection along the edge, by $\sim$17\% longer corner cells, or completely by an interleaved series--parallel connection of equal cells. Since the closed-form results are obtained instantly and the recurrent solution is fast for any device geometry and fluorophore parameters, the proposed approach provides a straightforward design tool for edge photovoltaic circuits of large-area LSCs.

\begin{backmatter}
\bmsection{Funding}
This research was funded by CETPartnership, the Clean Energy Transition Partnership under the 2023 joint call for research proposals, co-funded by the European Commission (GA N$^\circ$101069750) and with the funding organizations Swedish Energy Agency, Agencia Estatal de Investigaci\'on, and Ministero dell'Universit\`a e della Ricerca.

\bmsection{Acknowledgment}
The author acknowledges the KTH School of Engineering Sciences for support through its Sabbatical Program, and thanks Prof. Minoru Fujii and Prof. Hiroshi Sugimoto of Kobe University for hosting the sabbatical stay. The AI assistant Claude (Anthropic) was used for drafting and editing the text and for numerical checks; the author reviewed all content and takes full responsibility for it.

\bmsection{Disclosures}
The authors declare no conflicts of interest.

\bmsection{Data availability}
Data underlying the results presented in this paper are not publicly available at this time but may be obtained from the authors upon reasonable request.

\bmsection{Supplemental document}
See Supplement 1 for supporting content.
\end{backmatter}

\clearpage
\setcounter{section}{0}\setcounter{equation}{0}\setcounter{figure}{0}\setcounter{table}{0}
\renewcommand{\theequation}{S\arabic{equation}}\renewcommand{\thefigure}{S\arabic{figure}}\renewcommand{\thetable}{S\arabic{table}}\renewcommand{\thesection}{S\arabic{section}}
\renewcommand{\theHequation}{S\arabic{equation}}\renewcommand{\theHfigure}{S\arabic{figure}}\renewcommand{\theHtable}{S\arabic{table}}\renewcommand{\theHsection}{S\arabic{section}}
\makeatletter\renewcommand\@biblabel[1]{#1.}\makeatother
\begin{center}{\Large\bfseries Supplement 1}\\[6pt]{\large Edge photocurrent distribution in luminescent solar concentrators: analytical treatment and solar cell arrangement}\end{center}
\vspace{6pt}
\noindent Equations, figures and tables of the main text are referred to without the prefix ``S''.

\section{Path-length density at an edge point}\label{S-s:ny}
\paragraph{Geometry.} The rectangle occupies $0<x<L$, $0<y'<W$. All four edges are covered by solar cells and absorb every arriving photon (full edge coverage); the considered edge is at $x=0$. A point on this edge at height $y$ is at a distance $y$ from the corner $O=(0,0)$ and $W-y$ from the corner $(0,W)$. Directions into the device are parametrized by the angle $\beta\in(-\pi/2,\pi/2)$ from the inward normal, positive towards the corner $(0,W)$.

\paragraph{Response to a single emitter.} Let an isotropic emitter at point $P$ inside the device be at distance $r$ from the edge point, seen at angle $\beta$. In the plane of the slab, the fraction of its trapped emission that crosses the edge element $\dd y$ is the angle subtended by the projected element, $\cos\beta\,\dd y/r$, divided by $2\pi$:
\begin{equation}
\dd F=\frac{\cos\beta}{2\pi r}\,\dd y .
\label{S-eq:dF}
\end{equation}
Including attenuation, the edge response per unit length to a unit emitter at $P$ is
\begin{equation}
\Phi_y(P)=\frac{\cos\beta}{2\pi r}\,e^{-\alpha k r}.
\label{S-eq:Ey}
\end{equation}
Integration of $\Phi_y(P)$ along one edge gives the probability that a photon emitted at $P$ is collected by that edge, i.e. the edge collection function of Fujieda \emph{et al.}~\cite{S-Fujieda2019,S-Fujieda2020JPE} for a single wavelength and without the trapping factor.

\paragraph{Uniform emitters.} For emitters distributed uniformly with density $1/A$, the area element in polar coordinates centerd at the edge point is $r\,\dd r\,\dd\beta$, and the factor $r$ cancels against $1/r$ in Eq.~(\ref{S-eq:Ey}):
\begin{equation}
j(y)=\frac{1}{A}\int \Phi_y\,\dd A=\int_0^\infty n_y(r)\,e^{-\alpha kr}\,\dd r,\qquad
n_y(r)=\frac{1}{2\pi A}\int_{\Theta_y(r)}\cos\beta\,\dd\beta ,
\label{S-eq:jdef}
\end{equation}
where $\Theta_y(r)$ is the set of directions for which the point at distance $r$ lies inside the device. The arc of radius $r$ is cut by the side edge $y'=W$ when $r\sin\beta>W-y$, i.e. for $\sin\beta>(W-y)/r$, by the side edge $y'=0$ for $\sin\beta<-y/r$, and by the opposite edge when $r\cos\beta>L$, i.e. for $|\sin\beta|<\sigma(r)\equiv\sqrt{1-L^2/r^2}$ ($r>L$). Since $\int\cos\beta\,\dd\beta=\dd(\sin\beta)$, the integral over each half of the arc ($\beta>0$, towards the corner at a distance $W-y$, and $\beta<0$, towards the corner $O$ at a distance $y$) is simply the length of the allowed interval of $\sin\beta$. Thus,
\begin{equation}
n_y(r)=\frac{g_y(r)+g_{W-y}(r)}{2\pi A},
\label{S-eq:ny}
\end{equation}
where $g_s(r)$, $s=y$ or $W-y$, is the allowed range of $\sin\beta$ on the side of the corner at distance $s$ from the edge point. It is a piece-wise function of $r$. For $s\le L$ (the arc first reaches the side edge),
\begin{equation}
g_s(r)=\begin{cases}
1, & 0<r\le s,\\[2pt]
\dfrac{s}{r}, & s<r\le L,\\[6pt]
\dfrac{s}{r}-\sqrt{1-\dfrac{L^2}{r^2}}, & L<r\le\sqrt{s^2+L^2},\\[8pt]
0, & r>\sqrt{s^2+L^2},
\end{cases}
\label{S-eq:gs1}
\end{equation}
and for $s>L$ (the arc first reaches the opposite edge),
\begin{equation}
g_s(r)=\begin{cases}
1, & 0<r\le L,\\[2pt]
1-\sqrt{1-\dfrac{L^2}{r^2}}, & L<r\le s,\\[8pt]
\dfrac{s}{r}-\sqrt{1-\dfrac{L^2}{r^2}}, & s<r\le\sqrt{s^2+L^2},\\[8pt]
0, & r>\sqrt{s^2+L^2}.
\end{cases}
\label{S-eq:gs2}
\end{equation}
Here, $s/r$ is the upper limit of $\sin\beta$ set by the side edge, and $\sqrt{1-L^2/r^2}$ is the lower limit set by the opposite edge. Both forms are continuous and vanish at the distance to the corresponding opposite corner, $\sqrt{s^2+L^2}$. Equations~(\ref{S-eq:ny})--(\ref{S-eq:gs2}) are equivalent to the compact form $g_s(r)=\max[0,\min(1,s/r)-\sigma(r)]$, with $\sigma(r)=\sqrt{1-L^2/r^2}$ for $r>L$ and $\sigma(r)=0$ otherwise, which was used in the numerical implementation. For $r<\min(y,W-y,L)$ the full half-circle is inside the device, $g_y=g_{W-y}=1$, and $n_y=1/(\pi A)$. The support of $n_y$ ends at $r_\mathrm{max}=\sqrt{L^2+\max(y,W-y)^2}$, the distance to the farther opposite corner.

\paragraph{Sum rules.} (i) Integrating Eq.~(\ref{S-eq:ny}) over $r$ gives the loss-free profile $j_0(y)$, Eq.~(\ref{S-eq:j0}). (ii) Summing over the four edges and integrating along each edge,
\begin{equation}
\sum_\mathrm{edges}\int n_y(r)\,\dd y=p(r),
\label{S-eq:sum}
\end{equation}
where $p(r)$ is the in-plane path-length distribution of uniformly distributed isotropic emitters in the rectangle, normalized to $\int p\,\dd r=1$ \cite{S-Sychugov2019}. Equation~(\ref{S-eq:sum}) expresses the reciprocity between the emitter view (distance from a random interior point to the boundary along a random direction) and the receiver view used here (distance from a boundary point to the interior, weighted by $\cos\beta$). The same reciprocity underlies the invariance of the mean path length of light in bounded media \cite{S-Blanco2003,S-Davy2021}, observed experimentally in scattering media \cite{S-Savo2017}, and its evaluation for refractive objects of various shapes \cite{S-Majic2021}. At $r\to0$, Eq.~(\ref{S-eq:sum}) gives $p(0)=P/(\pi A)$, with $P$ the perimeter, valid for any convex shape; more generally $p(r)=(P/\pi A)[1-G(r)]$, where $G$ is the cumulative distribution of chord lengths for isotropic uniform (Cauchy--Crofton) chords. Without the $\cos\beta$ weight, $n_y$ reduces to the distribution of distances from a boundary point to uniformly distributed points, a quantity studied for random point sets and networks \cite{S-Tournus2011}. Both sum rules were verified numerically, and Eq.~(\ref{S-eq:ny}) was compared with Monte Carlo histograms of $1.2\times10^7$ in-plane paths [Fig.~\ref{fig:paths}(b)].

\section{Edge flux profile}\label{S-s:profile}
\paragraph{General form.} Interchanging the order of integration in Eq.~(\ref{S-eq:jdef}), with $R(\beta)$ the distance from the edge point to the boundary in direction $\beta$,
\begin{equation}
j(y)=\frac{1}{2\pi A}\int_{-\pi/2}^{\pi/2}\cos\beta\int_0^{R(\beta)}e^{-zr}\,\dd r\,\dd\beta
=\frac{1}{2\pi Az}\int_{-\pi/2}^{\pi/2}\cos\beta\,\big[1-e^{-zR(\beta)}\big]\dd\beta,\qquad z=\alpha k .
\label{S-eq:jR}
\end{equation}
The first term integrates to 2. Directions with $-\arctan(y/L)<\beta<\arctan[(W-y)/L]$ end on the opposite edge, $R=L/\cos\beta$; directions with $\beta>\arctan[(W-y)/L]$ end on the side edge $y'=W$, $R=(W-y)/\sin\beta$, and those with $\beta<-\arctan(y/L)$ end on the side edge $y'=0$, $R=y/|\sin\beta|$.

\paragraph{Loss-free profile.} For $z\to0$, Eq.~(\ref{S-eq:jR}) becomes $j_0=(2\pi A)^{-1}\int\cos\beta\,R(\beta)\,\dd\beta$, which is elementary:
\begin{equation}
j_0(y)=\frac{1}{2\pi A}\Big[L\Big(\arctan\frac{y}{L}+\arctan\frac{W-y}{L}\Big)+\frac{y}{2}\ln\Big(1+\frac{L^2}{y^2}\Big)+\frac{W-y}{2}\ln\Big(1+\frac{L^2}{(W-y)^2}\Big)\Big].
\label{S-eq:j0}
\end{equation}
For a square ($L=W$) the corner value is 0.654 of the center value.

\paragraph{Attenuated profile.} The opposite-edge contribution is
\begin{equation}
I_L=\int_{-\arctan(y/L)}^{\arctan[(W-y)/L]}\cos\beta\,e^{-zL/\cos\beta}\,\dd\beta ,
\label{S-eq:IL}
\end{equation}
an incomplete Bickley function. The Bickley functions (Bickley--Naylor functions)
\begin{equation}
\mathrm{Ki}_n(x)=\int_0^{\pi/2}\cos^{n-1}\theta\;e^{-x/\cos\theta}\,\dd\theta
\label{S-eq:Ki}
\end{equation}
are repeated integrals of the modified Bessel function $K_0$, with $\mathrm{Ki}_1(x)=\int_x^\infty K_0(t)\,\dd t$ and $\mathrm{Ki}_{n+1}'(x)=-\mathrm{Ki}_n(x)$. They describe Beer--Lambert attenuation averaged over oblique paths to a straight boundary and are commonly used in neutron and radiative transport in cylindrical and slab geometries. The incomplete functions $\mathrm{Ki}_n(x,\theta)$ have the upper limit $\theta<\pi/2$, so that $I_L=\mathrm{Ki}_2\big(zL,\arctan(y/L)\big)+\mathrm{Ki}_2\big(zL,\arctan[(W-y)/L]\big)$. Without attenuation, $\mathrm{Ki}_2(0,\theta)=\sin\theta$. For the side edge at a distance $s$ ($s=y$ or $W-y$), the substitution $c=\sin\beta$ gives
\begin{equation}
I_s=\int_{c_0}^{1}e^{-zs/c}\,\dd c=G_s(1)-G_s(c_0),\qquad c_0=\frac{s}{\sqrt{s^2+L^2}},\qquad
G_s(c)=c\,e^{-zs/c}-zs\,E_1(zs/c),
\label{S-eq:Is}
\end{equation}
since $\dd G_s/\dd c=e^{-zs/c}$ (using $E_1'(x)=-e^{-x}/x$). Collecting terms,
\begin{equation}
j(y)=\frac{1}{2\pi Az}\big[2-I_L-I_y-I_{W-y}\big],
\label{S-eq:jatt}
\end{equation}
which is Eq.~(\ref{eq:jatt}). Integrating $j(y)$ over the four edges recovers the waveguiding efficiency $f(\alpha)$ of the rectangle \cite{S-Sychugov2019}; this was checked numerically.

\paragraph{Limits.} For $\alpha L,\alpha W\gg1$ all terms $I$ vanish except near corners, and $j\to1/(\pi Az)$: only emitters within $\sim1/z$ of the edge contribute, and they see a half-plane. At the corner $O$ ($y=0$), directions with $\beta<0$ have $R=0$ and do not contribute, so $j\to1/(2\pi Az)$, one half of the edge-center value. The width of this corner dip is $\sim1/z$. Consequently, the corner-to-center contrast is non-monotonic in size: for a square of side $a$ it grows from its loss-free value for $\alpha a\lesssim3$, because attenuation favours emitters near the edge center, and decreases for larger devices as the dip narrows (Fig.~\ref{fig:size}).

\section{Out-of-plane path extension}\label{S-s:k}
A trapped photon travels at elevation angle $\psi$ relative to the plane of the slab, $|\psi|<\psi_m=\pi/2-\theta_c$, with $\sin\theta_c=1/n$. For isotropic emission the elevation angles are distributed with weight $\cos\psi$, and the true path is $l=r/\cos\psi$. The exact attenuation factor is
\begin{equation}
\big\langle e^{-\alpha r/\cos\psi}\big\rangle_\psi=\frac{\int_0^{\psi_m}\cos\psi\;e^{-\alpha r/\cos\psi}\,\dd\psi}{\sin\psi_m}.
\label{S-eq:exactpsi}
\end{equation}
To first order in $\alpha r$, this equals $e^{-\alpha kr}$ with
\begin{equation}
k=\Big\langle\frac{1}{\cos\psi}\Big\rangle=\frac{\int_0^{\psi_m}\dd\psi}{\int_0^{\psi_m}\cos\psi\,\dd\psi}=\frac{\psi_m}{\sin\psi_m}=1.128\quad(n=1.5),
\label{S-eq:k}
\end{equation}
the value obtained also by Soti \emph{et al.}~\cite{S-Soti1996}. A uniform weighting in $\psi$ gives 1.144, the value used in Ref.~\cite{S-Sychugov2019}. Values for other refractive indices: 1.060 ($n=1.2$), 1.181 (1.8), 1.265 (2.5), 1.337 (3.5). The difference between 1.128 and 1.144 changes waveguiding efficiencies by 1--2\%. If needed, $e^{-\alpha kr}$ in Eq.~(\ref{S-eq:jdef}) can be replaced by Eq.~(\ref{S-eq:exactpsi}) at the cost of one additional quadrature. The fraction of isotropic emission trapped by total internal reflection is $\delta=\cos\theta_c=0.745$.

The treatment is monochromatic, i.e. $\alpha$ is taken at a single emission wavelength or as an effective value. For emitters with spectral overlap between absorption and emission, self-absorption red-shifts the emission spectrum by an amount that depends on the propagation angle, so that the effective attenuation also depends on $\psi$. A spectrally resolved calculation applies Eqs.~(\ref{S-eq:jdef})--(\ref{S-eq:operator}) per wavelength. Measurements and modeling of the angular distribution of light exiting an LSC edge also show that the losses depend on the elevation angle \cite{S-Parel2015}; this dependence is captured by Eq.~(\ref{S-eq:exactpsi}) rather than by the single factor $k$.

\section{Recycling: scalar and operator series}\label{S-s:series}
\paragraph{Re-emission probability.} At a bulk event (rate $\alpha=\alpha_{sc}+\alpha_{re}+\alpha_{mx}$), the photon is scattered with probability $\alpha_{sc}/\alpha$, reabsorbed and re-emitted with probability $QY\,\alpha_{re}/\alpha$, or lost. Scattering and re-emission are taken as isotropic, so the new photon is trapped with probability $\delta$. The probability that an interaction produces a new trapped photon is $p=\delta(\alpha_{sc}+QY\alpha_{re})/\alpha$.

\paragraph{Scalar series.} A trapped photon emitted uniformly in the device reaches an edge with probability $f(\alpha)$ and interacts with probability $1-f$. If the interaction points (and hence the next-generation emitters) are again uniform, each generation contributes the same profile $j(y)$ multiplied by $[p(1-f)]^{n-1}$, which gives Eq.~(\ref{eq:scalar}). Integrated over the perimeter this reproduces the efficiency $f/[1-p(1-f)]$ of Refs.~\cite{S-Sychugov2019,S-Sychugov2020}.

\paragraph{Operator series.} In reality the interaction points are not uniform: photons emitted near an absorbing edge are collected before they interact, so secondary emitters are depleted near edges and corners. Let $s_n(P)$ be the density of emission points of generation $n$, $s_1=s_0=1/A$. The density of interaction events at $P$ per photon emitted at $P'$ is
\begin{equation}
K(P,P')=\frac{\alpha k\,e^{-\alpha k|P-P'|}}{2\pi|P-P'|},
\label{S-eq:K}
\end{equation}
(the in-plane path density $e^{-zr}/(2\pi r)$ per unit area, times the interaction rate per unit in-plane length, $z$). Then $s_{n+1}=pKs_n$, and the total emitter density $S=\sum_n s_n$ obeys the linear integral equation
\begin{equation}
S(P)=s_0+p\int_AK(P,P')\,S(P')\,\dd A' .
\label{S-eq:ie}
\end{equation}
The edge profile is the sum over generations of the edge response to each generation,
\begin{equation}
j_\mathrm{tot}(y)=\sum_{n\ge1}\int_A\Phi_y\,(pK)^{n-1}s_0\,\dd A=\int_A\Phi_y(P)\,S(P)\,\dd A,
\label{S-eq:operator}
\end{equation}
i.e. Eq.~(\ref{S-eq:jdef}) with the uniform emitter density replaced by $S$. All histories enter only through the position of the last emission. Replacing $K$ by its average over a uniform source, $\int_AK\,\dd A=1-f$, recovers the scalar series.

\paragraph{Numerical solution.} The device is divided into $n\times n$ square cells of size $h$ ($n=48$). Off-diagonal elements are $K_{ij}=z\,e^{-zD_{ij}}h^2/(2\pi D_{ij})$, with $D_{ij}$ the distance between cell centers. The self-interaction of a cell is approximated by that of a disk of equal area, $K_{ii}=1-\exp(-zh/\sqrt\pi)$. The edge response $\Phi_y$ is averaged over $6\times6$ sub-points in each source cell. The system $(I-pK)S=s_0$ is solved directly; the efficiency is obtained from the fraction of photons leaving each cell without interaction, $\sum_iS_i(1-\sum_jK_{ji})$.

\begin{figure}[htbp]
\centering\includegraphics[width=\textwidth]{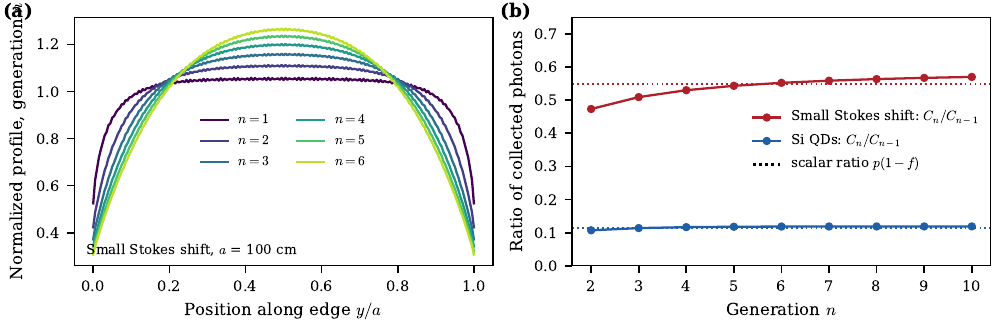}
\caption{(a) Normalized edge profiles of successive generations $n$ for the small-Stokes-shift emitter, $a=100$\,cm. (b) Ratio of photons collected in successive generations, compared with the scalar ratio $p(1-f)$ of Eq.~(\ref{eq:scalar}).}
\label{S-fig:gen}
\end{figure}

\paragraph{Convergence of the series.} The ratio of photons collected in successive generations, $C_n/C_{n-1}$, starts below the scalar value $p(1-f)$, because the second generation is depleted near the edges, and approaches the largest eigenvalue of $pK$, slightly above $p(1-f)$ (Fig.~\ref{S-fig:gen}). For the small-Stokes-shift example at $a=100$\,cm the ratio rises from 0.47 to 0.57 versus the scalar 0.55; for Si QDs from 0.107 to 0.119 versus 0.113. The profile of each generation is sharper than that of the previous one and converges to the fundamental mode of $K$. The series is therefore geometric only asymptotically.

\section{Monte Carlo simulation}\label{S-s:mc}
Emitter positions are sampled uniformly over the square and emission directions isotropically in three dimensions. Photons within the escape cones ($|\cos\theta|>\delta$, $\theta$ measured from the slab normal) are discarded; the efficiencies are therefore per trapped first emission. Each photon propagates in the plane along its projected direction. The free path to the next bulk event is drawn from an exponential distribution with mean $1/\alpha$ along the true path, so that the in-plane step is the true step times $\cos\psi$. If the in-plane distance to the boundary is shorter, the photon is collected and its arrival coordinate along the edge recorded. Otherwise the event is classified as matrix absorption (loss), scattering, or reabsorption with re-emission probability $QY$, in proportion to $\alpha_{mx}:\alpha_{sc}:\alpha_{re}$. New directions are isotropic, and new photons in the escape cones are lost. The generation index of each collected photon is stored. Arrival positions are pooled over the four edges (20 bins per edge, $1.5\times10^6$ photons per case). The slab thickness does not enter, since the path length of a trapped photon depends only on its in-plane displacement and elevation angle; reflections at the faces are taken as lossless.

The three parameter sets are listed in Table~\ref{S-tab:fl}. The Si QD parameters correspond to the glass/polymer laminates of Ref.~\cite{S-Zhou2026}; the other two sets represent a CuInS$_2$-type emitter and a small-Stokes-shift emitter of the CsPbBr$_3$/CdSe type. The waveguiding efficiencies obtained by the three methods for the profiles of Fig.~\ref{fig:prof} are compared in Table~\ref{S-tab:eff}.

\begin{table}[htbp]
\centering\small
\caption{Fluorophore parameters used in the simulations.}
\label{S-tab:fl}
\begin{tabular}{lccc}
\toprule
 & Si QDs & CuInS$_2$-type & small Stokes shift\\
\midrule
$\alpha_{mx}$, $\alpha_{re}$, $\alpha_{sc}$ (cm$^{-1}$) & 0.015, 0.007, 0.002 & 0.010, 0.020, 0.005 & 0.010, 0.080, 0\\
$QY$ & 0.55 & 0.90 & 0.95\\
\bottomrule
\end{tabular}
\end{table}

\begin{table}[htbp]
\centering\small
\caption{Waveguiding efficiency $\eta_\mathrm{wg}$ per trapped first emission of square LSCs with side $a$ for the fluorophores of Table~\ref{S-tab:fl}: Monte Carlo (MC), scalar series, Eq.~(\ref{eq:scalar}), and operator series, Eq.~(\ref{eq:operator}); the series columns are labeled by the equation numbers. The last row gives the share of collected photons re-emitted at least once.}
\label{S-tab:eff}
\setlength{\tabcolsep}{4pt}
\begin{tabular}{lccccccccc}
\toprule
 & \multicolumn{3}{c}{Si QDs} & \multicolumn{3}{c}{CuInS$_2$-type} & \multicolumn{3}{c}{small Stokes shift}\\
\cmidrule(lr){2-4}\cmidrule(lr){5-7}\cmidrule(lr){8-10}
 & MC & (\ref{eq:scalar}) & (\ref{eq:operator}) & MC & (\ref{eq:scalar}) & (\ref{eq:operator}) & MC & (\ref{eq:scalar}) & (\ref{eq:operator})\\
\midrule
$a=10$\,cm  & 0.902 & 0.902 & 0.902 & 0.909 & 0.909 & 0.908 & 0.832 & 0.832 & 0.830\\
$a=30$\,cm  & 0.742 & 0.742 & 0.741 & 0.752 & 0.753 & 0.750 & 0.577 & 0.586 & 0.573\\
$a=100$\,cm & 0.423 & 0.424 & 0.421 & 0.423 & 0.432 & 0.419 & 0.240 & 0.267 & 0.235\\
\midrule
recycled share, $a=100$\,cm & \multicolumn{3}{c}{11\%} & \multicolumn{3}{c}{33\%} & \multicolumn{3}{c}{50\%}\\
\bottomrule
\end{tabular}
\end{table}

For Si QDs, the scalar series reproduces the shape of the profile and the efficiency to within 0.3\% at all sizes; recycled photons constitute 2--11\% of the collected signal. For the CuInS$_2$-type emitter the scalar series overestimates the efficiency by 2\% at $a=100$\,cm, and for the small-Stokes-shift emitter by 1.6\% at 30\,cm and 11\% at 100\,cm, where half of the collected photons have been re-emitted at least once; in these cases the scalar profile is also too flat. The operator series follows the Monte Carlo profiles in all cases, with efficiencies within 1--2\%. The remaining difference is due to the approximation $l\approx kr$ (Section~\ref{S-s:k}) and to the source discretization. Cell-integrated quantities from the operator series (corner-to-center ratio for eight cells, series-to-parallel ratio for four cells) agree with Monte Carlo to within 0.01 for all fluorophores and sizes from 5 to 200\,cm (Fig.~\ref{fig:size}).

\section{Mirrors and partial edge coverage}\label{S-s:mirrors}
\paragraph{Unfolding.} A specular mirror at an edge is equivalent to an extension of the device by its mirror image \cite{S-Lu2023}; in-plane paths and their lengths are preserved, and the escape-cone geometry is unchanged.

\paragraph{Two PV edges opposite each other, two mirrors.} Unfolding across the mirrors gives an infinite strip of width $L$ with two absorbing edges, so the edge profile is exactly uniform. For an emitter at distance $x$ from one absorbing edge, the probability of reaching that edge is $\pi^{-1}\int_0^{\pi/2}e^{-zx/\cos\beta}\dd\beta=\pi^{-1}\mathrm{Ki}_1(zx)$, with $\mathrm{Ki}_n$ the Bickley functions of Eq.~(\ref{S-eq:Ki}). Averaging over $x$ and adding both edges,
\begin{equation}
f_\mathrm{s}(\alpha L)=\frac{2}{\pi z}\big[1-\mathrm{Ki}_2(z)\big]=\frac{2}{\pi z}-\frac{2}{\pi}\int_z^\infty\frac{K_1(t)}{t}\,\dd t,\qquad z=\alpha kL,
\label{S-eq:strip}
\end{equation}
where $K_1$ is the modified Bessel function. The two forms were checked numerically to agree to $10^{-9}$. Equation~(\ref{S-eq:strip}) is the exact form of the strip efficiency of Ref.~\cite{S-Lu2023}, whose truncated series deviates from it by at most 0.85\%. The edge profile is $j=f_\mathrm{s}/2W$.

\paragraph{One PV edge, three mirrors.} The unfolded device is a strip of width $2L$ with absorbing edges on both sides, again with a uniform profile.

\paragraph{Three PV edges, one mirror.} The unfolded device is a $2L\times W$ rectangle with four absorbing edges, so Eqs.~(\ref{S-eq:jatt})--(\ref{S-eq:operator}) apply directly. On the edges adjacent to the mirror, the profile is reduced only towards the corner shared with the other PV edge [Fig.~\ref{fig:design}(a)].

\paragraph{Small PV fraction.} If the PV covers a small fraction of an otherwise reflecting boundary, the trapped light approaches a uniform photon gas and the collected flux becomes independent of the position of the cell \cite{S-Song2018}. Partial edge reflection therefore flattens measured edge profiles.

\section{Rectangles and interleaved arrangements}\label{S-s:rect}
\paragraph{Interleaving.} For $N$ equal cells per edge with $N$ even, the profile is symmetric about the edge center, so cells $i$ and $N+1-i$ deliver the same current. Since $i$ and $N+1-i$ have opposite parity, the sum of currents of odd-position cells equals that of even-position cells exactly, for any profile shape, fluorophore or size. With four interleaved groups (positions modulo 4) this symmetry argument no longer pairs all cells, and a residual mismatch of 3--4\% remains (group current ratio 0.97 at $\alpha a=0.3$ and 0.96 at $\alpha a=3$).

\paragraph{Rectangles.} The current per unit edge length of the long edge relative to the short edge is 0.92--1.07 for aspect ratio 2 and 0.84--1.13 for aspect ratio 4 at $\alpha a_\mathrm{eq}=0$--3, with $a_\mathrm{eq}$ the side of the square of equal area. For a $2:1$ rectangle with sixteen equal cells on each long edge grouped into four interleaved blocks and eight cells on each short edge grouped into two blocks, the twelve blocks have currents within 2--4\% of each other.

\section{Other device shapes}\label{S-s:shapes}
\paragraph{Convex polygons.} The derivation of Section~\ref{S-s:profile} can be extended to any convex polygon with solar cells on all edges, including the regular shapes considered in Ref.~\cite{S-Sychugov2020}. For a point $P$ on the solar cell edge, every other edge $i$ is a straight line at a perpendicular distance $d_i$ from $P$. Its outward normal is seen at an angle $\phi_i$ from the inward normal at $P$, and the edge itself is seen within the angles $\beta_{a,i}<\beta<\beta_{b,i}$, where the distance to the boundary is $R=d_i/\cos(\beta-\phi_i)$. With $u=\beta-\phi_i$ and $\cos\beta=\cos u\cos\phi_i-\sin u\sin\phi_i$, the contribution of edge $i$ to Eq.~(\ref{S-eq:jR}) separates into an incomplete Bickley function and a closed-form term:
\begin{equation}
\int_{\beta_{a,i}}^{\beta_{b,i}}\!\cos\beta\,e^{-zR}\,\dd\beta=\cos\phi_i\,T_{1,i}-\sin\phi_i\,T_{2,i},
\label{S-eq:poly1}
\end{equation}
\begin{equation}
T_{1,i}=\int_{u_{a,i}}^{u_{b,i}}\!\cos u\,e^{-zd_i/\cos u}\,\dd u,\qquad T_{2,i}=G_i(\cos u_{a,i})-G_i(\cos u_{b,i}),
\label{S-eq:poly2}
\end{equation}
where $G_i(c)=c\,e^{-zd_i/c}-zd_i\,E_1(zd_i/c)$ is the function of Eq.~(\ref{S-eq:Is}) with $s=d_i$, since $\dd[-G_i(\cos u)]/\dd u=\sin u\,e^{-zd_i/\cos u}$. The edge profile is then
\begin{equation}
j(y)=\frac{1}{2\pi Az}\sum_i\Big[\sin\beta_{b,i}-\sin\beta_{a,i}-\cos\phi_i\,T_{1,i}+\sin\phi_i\,T_{2,i}\Big],
\label{S-eq:jpoly}
\end{equation}
where the sum runs over all edges except the one containing $P$. For a rectangle, the opposite edge has $\phi=0$ (pure Bickley term) and the side edges have $\phi=\pm\pi/2$ (pure $E_1$ terms), which gives Eq.~(\ref{S-eq:jatt}). Without attenuation the result is elementary,
\begin{equation}
j_0(y)=\frac{1}{2\pi A}\sum_i d_i\Big[\cos\phi_i\,(u_{b,i}-u_{a,i})+\sin\phi_i\ln\frac{\cos u_{b,i}}{\cos u_{a,i}}\Big],
\label{S-eq:j0poly}
\end{equation}
and the path density at the edge point is, as in Eq.~(\ref{S-eq:ny}), the length of the allowed interval of $\sin\beta$:
\begin{equation}
n_y(r)=\frac{1}{2\pi A}\sum_i\Big|\big\{\beta\in(\beta_{a,i},\beta_{b,i}):\ \cos(\beta-\phi_i)<d_i/r\big\}\Big|_{\sin\beta},
\label{S-eq:nypoly}
\end{equation}
where the whole interval contributes for $r\le d_i$, and the directions with $|\beta-\phi_i|<\arccos(d_i/r)$ are excluded for $r>d_i$. Equations~(\ref{S-eq:jpoly})--(\ref{S-eq:nypoly}) reproduce Eqs.~(\ref{S-eq:ny})--(\ref{S-eq:jatt}) for a square, and $\int n_y\,\dd r=j_0$ holds for all shapes. The recycling treatment of Section~\ref{S-s:series} applies without changes, with $\Phi_y$ and $K$ evaluated in the given shape.

\paragraph{Corner limit.} For $\alpha a\gg1$ the profile is flat, $j\to1/(\pi Az)$, except near the corners. At a corner with interior angle $\theta$, only the directions within this angle contribute, which gives $\int\cos\beta\,\dd\beta=1-\cos\theta$ instead of 2, i.e.
\begin{equation}
j_\mathrm{corner}/j_\infty=\sin^2(\theta/2),
\label{S-eq:corner}
\end{equation}
which equals 1/4 for an equilateral triangle, 1/2 for a square, 3/4 for a regular hexagon, and approaches 1 for a circle.

\begin{figure}[htbp]
\centering\includegraphics[width=\textwidth]{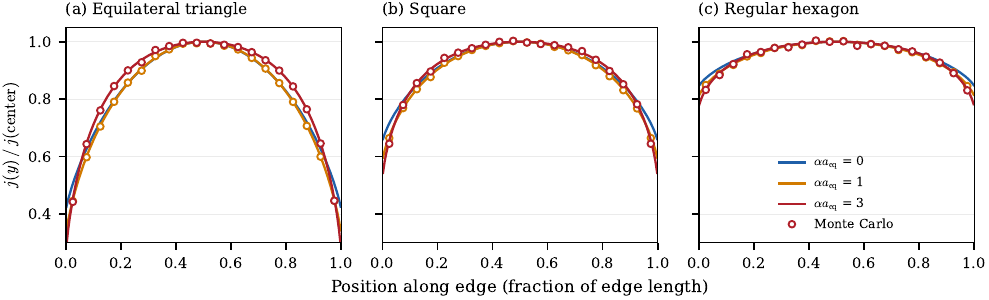}
\caption{Normalized edge profiles of (a) an equilateral triangle, (b) a square and (c) a regular hexagon of equal area for $\alpha a_\mathrm{eq}=0$, 1 and 3. Lines are from Eq.~(\ref{S-eq:jpoly}), and symbols are from Monte Carlo simulations of in-plane paths for $\alpha a_\mathrm{eq}=1$ and 3.}
\label{S-fig:shapes}
\end{figure}

\paragraph{Regular shapes.} Edge profiles of an equilateral triangle, a square and a regular hexagon of equal area are shown in Fig.~\ref{S-fig:shapes}, and the corresponding design quantities are listed in Table~\ref{S-tab:shapes} ($a_\mathrm{eq}=\sqrt{A}$). Monte Carlo simulations of in-plane paths ($2\times10^6$ emitters per shape, uniform in-plane directions, survival probability $e^{-zR}$) agree with Eq.~(\ref{S-eq:jpoly}) within the statistical error ($<$1\% for bin-averaged values), and the integral of $j$ over the perimeter reproduces the simulated waveguiding efficiency within 0.2\%. In agreement with Ref.~\cite{S-Sychugov2020}, all regular shapes have a similar waveguiding efficiency for the same area. The edge uniformity, however, strongly depends on the shape. The corner-to-center ratio is 0.27--0.42 for a triangle, 0.52--0.65 for a square and 0.77--0.85 for a hexagon, and the current of four equal series-connected cells per edge is reduced by 15--17\%, 8--9\% and 3--4\%, respectively. Accordingly, equal-current corner cells should be $\sim$35\% longer than the central ones for a triangle, $\sim$17\% for a square and $\sim$8\% for a hexagon. For a circle, all edge points are equivalent, the profile is uniform, $j=f/(2\pi R_c)$ with $f$ from Ref.~\cite{S-Sychugov2020}, and no mismatch occurs. Since the edge profiles of all regular polygons are symmetric, the interleaved series--parallel connection (Section~\ref{S-s:rect}) gives exactly matched blocks for any regular shape.

\begin{table}[htbp]
\centering\small
\caption{Edge profile characteristics of regular LSCs of equal area: waveguiding efficiency per trapped photon $f$, corner-to-center flux ratio, edge-average-to-center ratio, current of four equal series-connected cells per edge relative to parallel connection, and equal-current lengths of four cells per edge (fraction of the edge length, corner and central cells).}
\label{S-tab:shapes}
\begin{tabular}{llccccc}
\toprule
Shape & $\alpha a_\mathrm{eq}$ & $f$ & corner/center & average/center & 4 cells, series & equal-current lengths\\
\midrule
Triangle & 0 & 1 & 0.415 & 0.830 & 0.843 & 0.286, 0.214\\
         & 1 & 0.633 & 0.328 & 0.820 & 0.828 & 0.289, 0.211\\
         & 3 & 0.338 & 0.266 & 0.847 & 0.848 & 0.284, 0.216\\
\midrule
Square   & 0 & 1 & 0.654 & 0.907 & 0.919 & 0.269, 0.231\\
         & 1 & 0.621 & 0.586 & 0.897 & 0.908 & 0.272, 0.228\\
         & 3 & 0.317 & 0.524 & 0.905 & 0.912 & 0.271, 0.229\\
\midrule
Hexagon  & 0 & 1 & 0.845 & 0.958 & 0.966 & 0.258, 0.242\\
         & 1 & 0.616 & 0.811 & 0.952 & 0.960 & 0.260, 0.240\\
         & 3 & 0.309 & 0.774 & 0.952 & 0.958 & 0.260, 0.240\\
\midrule
Circle   & -- & -- & 1 & 1 & 1 & 0.25, 0.25\\
\bottomrule
\end{tabular}
\end{table}


\begin{thebibliography}{99}
\bibitem{Debije2012} M. G. Debije and P. P. C. Verbunt, ``Thirty years of luminescent solar concentrator research: solar energy for the built environment,'' Adv. Energy Mater. \textbf{2}, 12--35 (2012).
\bibitem{Meinardi2017NRM} F. Meinardi, F. Bruni, and S. Brovelli, ``Luminescent solar concentrators for building-integrated photovoltaics,'' Nat. Rev. Mater. \textbf{2}, 17072 (2017).
\bibitem{Richards2023} B. S. Richards and I. A. Howard, ``Luminescent solar concentrators for building integrated photovoltaics: opportunities and challenges,'' Energy Environ. Sci. \textbf{16}, 3214--3239 (2023).
\bibitem{BorjaBlock2024} A. Borja Block \emph{et al.}, ``Colouring solutions for building integrated photovoltaic modules: a review,'' Energy Build. \textbf{314}, 114253 (2024).
\bibitem{Magrini2020} A. Magrini \emph{et al.}, ``From nearly zero energy buildings (NZEB) to positive energy buildings (PEB): the next challenge,'' Dev. Built Environ. \textbf{3}, 100019 (2020).
\bibitem{Meinardi2024} F. Meinardi \emph{et al.}, ``Certification grade quantum dot luminescent solar concentrator glazing with optical wireless communication capability for connected sustainable architecture,'' Adv. Energy Mater., 2304006 (2024).
\bibitem{Traverse2017} C. J. Traverse \emph{et al.}, ``Emergence of highly transparent photovoltaics for distributed applications,'' Nat. Energy \textbf{2}, 849--860 (2017).
\bibitem{Li2023NRM} Y. Li \emph{et al.}, ``Semitransparent organic photovoltaics for building-integrated photovoltaic applications,'' Nat. Rev. Mater. \textbf{8}, 186--201 (2023).
\bibitem{Li2016} H. Li \emph{et al.}, ``Doctor-blade deposition of quantum dots onto standard window glass for low-loss large-area luminescent solar concentrators,'' Nat. Energy \textbf{1}, 16157 (2016).
\bibitem{Zhou2026} J. Zhou \emph{et al.}, ``Large-area luminescent solar concentrators with high optical clarity towards integration with smart windows,'' Sol. Energy Mater. Sol. Cells \textbf{302}, 114359 (2026).
\bibitem{Bergren2018} M. R. Bergren \emph{et al.}, ``High-performance CuInS$_2$ quantum dot laminated glass luminescent solar concentrators for windows,'' ACS Energy Lett. \textbf{3}, 520--525 (2018).
\bibitem{Huang2022} J. Huang \emph{et al.}, ``Large-area transparent `quantum dot glass' for building-integrated photovoltaics,'' ACS Photonics \textbf{9}, 2499--2509 (2022).
\bibitem{Aste2019} N. Aste \emph{et al.}, ``Triggering a large scale luminescent solar concentrators market: the smart window project,'' J. Clean. Prod. \textbf{219}, 35--45 (2019).
\bibitem{Yang2022} C. Yang \emph{et al.}, ``Consensus statement: standardized reporting of power-producing luminescent solar concentrator performance,'' Joule \textbf{6}, 8--15 (2022).
\bibitem{deBruin2024} T. A. de Bruin and W. G. J. H. M. van Sark, ``Shining lights on limits: optimizing luminescent solar concentrators for solar windows,'' Adv. Photonics Res., 2400068 (2024).
\bibitem{Needell2021} D. R. Needell \emph{et al.}, ``Solar power windows: connecting scientific advances to market signals,'' Energy \textbf{219}, 119567 (2021).
\bibitem{Ravyts2019} S. Ravyts \emph{et al.}, ``Review on building-integrated photovoltaics electrical system requirements and module-integrated converter recommendations,'' Energies \textbf{12}, 1532 (2019).
\bibitem{Heidler1981} K. Heidler, ``Efficiency and concentration ratio measurements of fluorescent solar concentrators using a xenon measurement system,'' Appl. Opt. \textbf{20}, 773--777 (1981).
\bibitem{Decardona1985} M. S. de Cardona \emph{et al.}, ``Edge effect on luminescent solar concentrators,'' Solar Cells \textbf{15}, 225--230 (1985).
\bibitem{Loh1986} E. Loh and D. J. Scalapino, ``Luminescent solar concentrators -- effects of shape on efficiency,'' Appl. Opt. \textbf{25}, 1901--1907 (1986).
\bibitem{Wilson2017} L. R. Wilson, E. Klampaftis, and B. S. Richards, ``Enhancement of power output from a large-area luminescent solar concentrator with 4.8$\times$ concentration via solar cell current matching,'' IEEE J. Photovolt. \textbf{7}, 802--809 (2017).
\bibitem{WilsonThesis} L. R. Wilson, ``Luminescent solar concentrators: a study of optical properties, re-absorption and device optimisation,'' Ph.D. thesis, Heriot-Watt University (2010).
\bibitem{Han2022} S. Han \emph{et al.}, ``Luminescence-guided and visibly transparent solar concentrators based on silicon quantum dots,'' Opt. Express \textbf{30}, 26896--26911 (2022).
\bibitem{Li2020} Y. Li, Y. Sun, and Y. Zhang, ``Regional measurements to analyze large-area luminescent solar concentrators,'' Renew. Energy \textbf{160}, 127--135 (2020).
\bibitem{Zhou2022} J. Zhou, J. Huang, and I. Sychugov, ``Optical center of a luminescent solar concentrator,'' Opt. Lett. \textbf{47}, 4985 (2022).
\bibitem{Sychugov2019} I. Sychugov, ``Analytical description of a luminescent solar concentrator,'' Optica \textbf{6}, 1046--1049 (2019).
\bibitem{Sychugov2020} I. Sychugov, ``Geometry effects on luminescence solar concentrator efficiency: analytical treatment,'' Appl. Opt. \textbf{59}, 5715 (2020).
\bibitem{Zheng2022} Z. Zheng \emph{et al.}, ``Modeling and comparison of bulk and thin-film luminescent solar concentrators based on colloidal perovskite quantum dots,'' Opt. Lett. \textbf{47}, 4367--4370 (2022).
\bibitem{deBruin2023} T. A. de Bruin and W. G. J. H. M. van Sark, ``Numerical method for calculation of power conversion efficiency and colorimetrics of rectangular luminescent solar concentrators,'' Sol. RRL \textbf{7}, 2200787 (2023).
\bibitem{tenKate2014} O. M. ten Kate, K. M. Hooning, and E. van der Kolk, ``Quantifying self-absorption losses in luminescent solar concentrators,'' Appl. Opt. \textbf{53}, 5238--5245 (2014).
\bibitem{Merkx2017} E. P. J. Merkx, O. M. ten Kate, and E. van der Kolk, ``Rapid optimization of large-scale luminescent solar concentrators: evaluation for adoption in the built environment,'' Opt. Express \textbf{25}, A547--A563 (2017).
\bibitem{Matsumura2020} R. Matsumura, Y. Tsutsumi, and I. Fujieda, ``Concentric re-emission pattern from a planar waveguide with a thin uniform luminescent layer,'' Appl. Opt. \textbf{59}, 1703--1710 (2020).
\bibitem{Klimov2016} V. I. Klimov \emph{et al.}, ``Quality factor of luminescent solar concentrators and practical concentration limits attainable with semiconductor quantum dots,'' ACS Photonics \textbf{3}, 1138--1148 (2016).
\bibitem{Fujieda2019} I. Fujieda \emph{et al.}, ``Edge collection function: an analytical expression for the optical efficiency of an energy-harvesting device based on photoluminescence,'' Opt. Eng. \textbf{58}, 104101 (2019).
\bibitem{Fujieda2020JPE} I. Fujieda and Y. Tsutsumi, ``Extending the concept of edge collection function to polygonal and curved planar luminescent waveguides,'' J. Photonics Energy \textbf{10}, 044501 (2020).
\bibitem{Yunoki2019} K. Yunoki \emph{et al.}, ``Cross talk and optical efficiency of an energy-harvesting color projector utilizing ceramic phosphors,'' Appl. Opt. \textbf{58}, 9896--9903 (2019).
\bibitem{Soti1996} R. Soti \emph{et al.}, ``Photon transport in luminescent solar concentrators,'' J. Lumin. \textbf{68}, 105--114 (1996).
\bibitem{Goetzberger1977} A. Goetzberger and W. Greubel, ``Solar energy conversion with fluorescent collectors,'' Appl. Phys. \textbf{14}, 123--139 (1977).
\bibitem{Zhang2015} J. Zhang \emph{et al.}, ``Optimization of large-size glass laminated luminescent solar concentrators,'' Sol. Energy \textbf{117}, 260--267 (2015).
\bibitem{Lu2023} X. Lu \emph{et al.}, ``Luminescent solar concentrator efficiency versus edge solar cell coverage,'' Opt. Lett. \textbf{48}, 4197 (2023).
\bibitem{Song2018} H.-J. Song \emph{et al.}, ``Performance limits of luminescent solar concentrators tested with seed/quantum-well quantum dots in a selective-reflector-based optical cavity,'' Nano Lett. \textbf{18}, 395--404 (2018).
\bibitem{Parel2015} T. S. Parel \emph{et al.}, ``Modelling and experimental analysis of the angular distribution of the emitted light from the edge of luminescent solar concentrators,'' Opt. Mater. \textbf{42}, 532--537 (2015).
\end{thebibliography}

\begin{thebibliography}{S99}
\bibitem[S1]{S-Sychugov2019} I. Sychugov, ``Analytical description of a luminescent solar concentrator,'' Optica \textbf{6}, 1046--1049 (2019).
\bibitem[S2]{S-Sychugov2020} I. Sychugov, ``Geometry effects on luminescence solar concentrator efficiency: analytical treatment,'' Appl. Opt. \textbf{59}, 5715 (2020).
\bibitem[S3]{S-Fujieda2019} I. Fujieda \emph{et al.}, ``Edge collection function: an analytical expression for the optical efficiency of an energy-harvesting device based on photoluminescence,'' Opt. Eng. \textbf{58}, 104101 (2019).
\bibitem[S4]{S-Fujieda2020JPE} I. Fujieda and Y. Tsutsumi, ``Extending the concept of edge collection function to polygonal and curved planar luminescent waveguides,'' J. Photonics Energy \textbf{10}, 044501 (2020).
\bibitem[S5]{S-Blanco2003} S. Blanco and R. Fournier, ``An invariance property of diffusive random walks,'' Europhys. Lett. \textbf{61}, 168 (2003).
\bibitem[S6]{S-Davy2021} M. Davy \emph{et al.}, ``Mean path length invariance in wave-scattering beyond the diffusive regime,'' Commun. Phys. \textbf{4}, 85 (2021).
\bibitem[S7]{S-Savo2017} R. Savo \emph{et al.}, ``Observation of mean path length invariance in light-scattering media,'' Science \textbf{358}, 765--768 (2017).
\bibitem[S8]{S-Parel2015} T. S. Parel \emph{et al.}, ``Modelling and experimental analysis of the angular distribution of the emitted light from the edge of luminescent solar concentrators,'' Opt. Mater. \textbf{42}, 532--537 (2015).
\bibitem[S9]{S-Majic2021} M. Majic, W. R. C. Somerville, and E. C. Le Ru, ``Mean path length inside nonscattering refractive objects,'' Phys. Rev. A \textbf{103}, L031502 (2021).
\bibitem[S10]{S-Tournus2011} F. Tournus, ``Random nanoparticle deposition: inter-particle distances in 2D, 3D, and multilayer samples,'' J. Nanopart. Res. \textbf{13}, 5211--5223 (2011).
\bibitem[S11]{S-Soti1996} R. Soti \emph{et al.}, ``Photon transport in luminescent solar concentrators,'' J. Lumin. \textbf{68}, 105--114 (1996).
\bibitem[S12]{S-Zhou2026} J. Zhou \emph{et al.}, ``Large-area luminescent solar concentrators with high optical clarity towards integration with smart windows,'' Sol. Energy Mater. Sol. Cells \textbf{302}, 114359 (2026).
\bibitem[S13]{S-Lu2023} X. Lu \emph{et al.}, ``Luminescent solar concentrator efficiency versus edge solar cell coverage,'' Opt. Lett. \textbf{48}, 4197 (2023).
\bibitem[S14]{S-Song2018} H.-J. Song \emph{et al.}, ``Performance limits of luminescent solar concentrators tested with seed/quantum-well quantum dots in a selective-reflector-based optical cavity,'' Nano Lett. \textbf{18}, 395--404 (2018).
\end{thebibliography}
\end{document}